\documentclass[12pt,a4paper]{iopart}
\usepackage{setstack,iopams}
\usepackage{graphicx} % for figures
\usepackage[caption=false]{subfig}

\newcommand{\ket}[1]{{\left| #1 \right\rangle}}
\newcommand{\ketbra}[2]{{\left| #1 \middle\rangle \middle \langle #2 \right|}}

\begin{document}

\title{Quantum Walk Search through Potential Barriers}

\author{Thomas G Wong\footnote{Present address: Department of Computer Science, University of Texas at Austin, Austin, TX 78712}}
\address{Faculty of Computing, University of Latvia, Rai\c{n}a bulv.~19, R\=\i ga, LV-1586, Latvia}
\ead{\mailto{twong@cs.utexas.edu}}

\begin{abstract}
	An ideal quantum walk transitions from one vertex to another with perfect fidelity, but in physical systems, the particle may be hindered by potential energy barriers. Then the particle has some amplitude of tunneling through the barriers, and some amplitude of staying put. We investigate the algorithmic consequence of such barriers for the quantum walk formulation of Grover's algorithm. We prove that the failure amplitude must scale as $O(1/\sqrt{N})$ for search to retain its quantum $O(\sqrt{N})$ runtime; otherwise, it searches in classical $O(N)$ time. Thus searching larger ``databases'' requires increasingly reliable hop operations or error correction. This condition holds for both discrete- and continuous-time quantum walks.
\end{abstract}


%-------------------------------------------------------------------------------
% Main Matter
%-------------------------------------------------------------------------------

\section{Introduction}

Classical random walks, or Markov chains, serve as the foundation of many classical probabilistic algorithms \cite{Norris1998}. It is no surprise, then, that their quantum analogues, quantum walks \cite{ADZ1993,Ambainis2003,Kempe2003}, similarly form the basis of many important quantum algorithms. This includes quantum search \cite{SKW2003}, element distinctness \cite{Ambainis2004}, and evaluating NAND trees \cite{FGG2008}, all of which provably yield polynomial speedups over the best possible classical algorithms. In other applications \cite{CCDFGS2003,CSV2007}, the speedup can even be exponential, and any quantum algorithm can be efficiently simulated by a quantum walk \cite{CGW2013}.

These algorithms assume that the quantum particle transitions from one vertex of a graph to another without hindrance. In physical systems, however, this may not be the case. As explained in \cite{Wong17}, various physical processes can prevent the particle from hopping, and the obstacles can be modeled as potential energy barriers. Then the randomly walking quantum particle must tunnel through the barriers in order to transition between vertices. As shown in \cite{Wong17}, even with the barriers, a quantum walk on the one-dimensional line still exhibits the ballistic propagation that is characteristic of quantum walks \cite{Ambainis2001}. Despite this, the work leaves open whether the barriers have \emph{algorithmic} consequences.

In this paper, we investigate whether potential barriers have algorithmic consequences by exploring their effect on the discrete- and continuous-time quantum walk formulations of Grover's algorithm \cite{Grover1996}. In the next section, we introduce discrete-time quantum walks and apply them to the unstructured search problem with potential barriers. We prove that, for the search to keep its quantum $O(\sqrt{N})$ runtime, the amplitude of failing to hop must go to zero sufficiently quickly, specifically as $O(1/\sqrt{N})$. Otherwise, the algorithm searches in $O(N)$ time, losing its speedup over classical. Thus the potential barriers have significant algorithmic consequences; either the potential barriers must be asymptotically eliminated, or error correction \cite{Wong15} must be utilized. Following, we analyze search with continuous-time quantum walks with potential barriers, arriving at the same condition for the error amplitude. Finally, we end with concluding remarks.

%-------------------------------------------------------------------------------
% Section
%-------------------------------------------------------------------------------

\section{Discrete-Time Quantum Walks}

\begin{figure}
\begin{center}
	\includegraphics{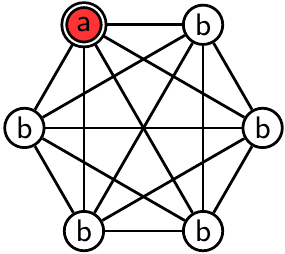}
	\caption{\label{fig:complete} The complete graph with $N = 6$ vertices. A vertex is marked, as indicated by a double circle. Identically evolving vertices are identically colored and labeled.}
\end{center}
\end{figure}

Grover's algorithm solves the unstructured search problem, which is equivalent to searching the complete graph of $N$ vertices for a marked vertex \cite{AKR2005,CG2004}, as illustrated in \fref{fig:complete}. First let us introduce the quantum walk without the searching component. The vertices of the graph label computational basis states $\{ \ket{1}, \ket{2}, \dots, \ket{N} \}$ of an $N$-dimensional ``vertex'' Hilbert space $\mathbb{C}^N$. In discrete-time, these vertices only support a trivial quantum walk \cite{Meyer1996a,Meyer1996b}, so assuming the graph is $d$-regular, we define an additional $d$-dimensional ``coin'' Hilbert space $\mathbb{C}^d$ spanned by the $d$ directions in which the particle can hop from one vertex to another. So the system evolves in $\mathbb{C}^N \otimes \mathbb{C}^d$. For the complete graph, each vertex is connected to the $N-1$ other vertices, so $d = N-1$. 

Without potential barriers, each step of the quantum walk is performed by applying a ``coin flip'' followed by a hop/shift:
\[ U_0 = S \cdot ( I_N \otimes C_0 ), \]
where $C_0$ is the ``Grover diffusion'' coin \cite{SKW2003},
\[ C_0 = 2 \ketbra{s_c}{s_c} - I_d, \]
where $\ket{s_c} = \sum_{i=1}^d \ket{i} / \sqrt{d}$ is the equal superposition over the coin space, and $S$ is the flip-flop shift \cite{AKR2005} that causes the particle to jump from one vertex to another and then turn around (\textit{e.g.}, $S \ket{1} \otimes \ket{1 \to 2} = \ket{2} \otimes \ket{2 \to 1}$).

Now with potential barriers, the particle has some amplitude $\alpha$ of successfully hopping and some (related) amplitude $\beta$ of staying put, leading to the following faulty shift:
\[ S \to \alpha S + \beta I. \]
Since $S$ is both unitary and Hermitian, $\alpha S + \beta I$ is unitary provided $|\alpha|^2 + |\beta|^2 = 1$ and $\alpha\beta^* + \beta\alpha^* = 0$. Given these constraints, we parameterize
\[ \alpha = \cos\phi \quad {\rm and} \quad \beta = \rmi \sin\phi. \]
Then the quantum walk operator is
\[ U_0' = (\cos(\phi) \, S + \rmi\sin(\phi) \, I) \cdot (I_N \otimes C_0). \]
Note that $\phi$ can be interpreted as a failure parameter. When $\phi = 0$, the walk is ideal and there are no potential barriers. But as $\phi \to \pi/2$, the barriers get worse and worse, causing the hop to fail more.

\begin{figure}
\begin{center}
	\includegraphics{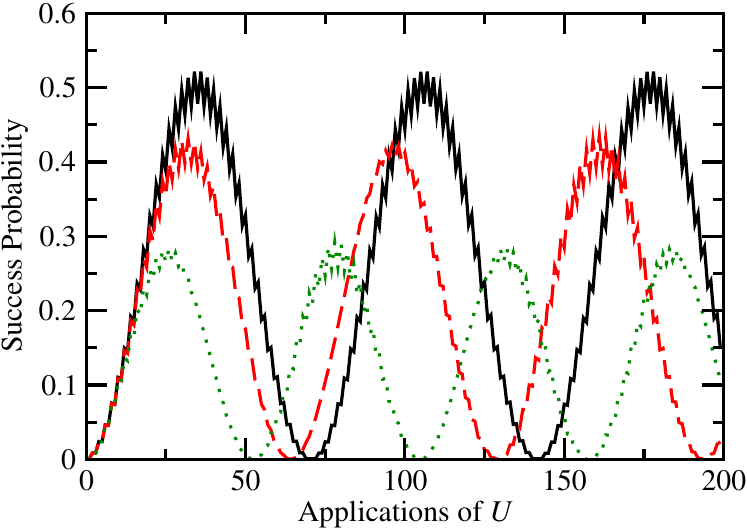}
	\caption{\label{fig:discrete_N1024} Success probability as a function of the number of applications of $U$ for search on the complete graph with $N=1024$ vertices with $\phi = 0$ (no potential barriers), $\phi = 0.02$, and $\phi = 0.04$ corresponding to the solid black, dashed red, and dotted green curves, respectively.}
\end{center}
\end{figure}

If the particle begins in the uniform distribution, \textit{i.e.}, $\ket{\psi_0} = \ket{s_v} \otimes \ket{s_c}$, where $\ket{s_v} = \sum_{i=1}^N \ket{i} / \sqrt{N}$ is the equal superposition over the vertex space, then the potential barriers makes no difference, \textit{i.e.}, $U_0 \ket{\psi_0} = \ket{\psi_0}$ and $U_0' \ket{\psi_0} = \ket{\psi_0}$. To turn this quantum walk into a search problem, we look for a particular ``marked'' vertex $\ket{a}$, as shown in \fref{fig:complete}, by querying an oracle $R_a$ that flips the phase of the marked vertex, \textit{i.e.}, $R_a \ket{a} = -\ket{a}$ and $R_a \ket{x} = \ket{x}, \forall x \ne a$. Querying the oracle with each step of the quantum walk \cite{Wong11}, the search is performed by repeatedly applying
\begin{equation}
	\label{eq:U}
	U = (\cos(\phi) \, S + \rmi\sin(\phi) \, I) \cdot (I_N \otimes C_0) \cdot (R_a \otimes I_d)
\end{equation}
to the initial equal superposition $\ket{\psi_0} = \ket{s_v} \otimes \ket{s_c}$. As proved in \cite{Wong10}, when the particle freely transitions from one vertex to another without the potential barriers (\textit{i.e.}, $\phi = 0$), the success probability reaches $1/2$ after $\pi\sqrt{N} / 2\sqrt{2}$ applications of $U$, as shown in \fref{fig:discrete_N1024}, which results in a $\Theta(\sqrt{N})$ search algorithm with the expected constant number of classical repetitions to boost the success probability near $1$. As in typical discrete-time quantum walk algorithms, this success probability of $1/2$ assumes that only the position of the particle is measured. If the internal state of the particle is additionally measured, however, the success probability can be boosted to $1$ \cite{Wong18}.

With the potential barriers (\textit{i.e.}, $\phi > 0$), the particle has some amplitude of not hopping. Even for small values of $\phi$, this can significantly impair the search, also shown in \fref{fig:discrete_N1024}. Let us find how large $\phi$ can be such that we still search in $\Theta(\sqrt{N})$ time. In doing so, we will get a sense for how small the potential barriers must be for quantum walks to search quickly.

To do this, we explicitly work out the evolution of the system. As shown in \fref{fig:complete}, there are only two types of vertices: the marked red $a$ vertex and the unmarked white $b$ vertices. By symmetry, the $b$ vertices evolve identically. Now the $a$ vertex can only point towards $b$ vertices, while $b$ vertices can either point towards the $a$ vertex or other $b$ vertices. Thus the system evolves in a 3D subspace spanned by these vertices and directions:
\begin{eqnarray*}
	\ket{ab} = \ket{a} \otimes \frac{1}{\sqrt{N-1}} \sum_b \ket{a \to b}, \\
	\ket{ba} = \frac{1}{\sqrt{N-1}} \sum_b \ket{b} \otimes \ket{b \to a}, \\
	\ket{bb} = \frac{1}{\sqrt{N-1}} \sum_b \ket{b} \otimes \frac{1}{\sqrt{N-2}} \sum_{b'} \ket{b \to b'}.
\end{eqnarray*}
In this $\{ \ket{ab}, \ket{ba}, \ket{bb} \}$ basis, initial equal superposition state is
\[ \ket{\psi_0} = \frac{1}{\sqrt{N}} \left( \!\! \begin{array}{c}
	1 \\
	1 \\
	\sqrt{N-2}
\end{array} \!\! \right), \]
and the operators that make up the search operator $U$ \eref{eq:U} are
\begin{eqnarray*}
	S = \left( \!\! \begin{array}{ccc}
		0 & 1 & 0  \\
		1 & 0 & 0 \\
		0 & 0 & 1 \\
	\end{array} \!\! \right), \quad
	(I_N \otimes C_0) = \left( \!\! \begin{array}{ccc}
		1 & 0 & 0 \\
		0 & -\cos\theta & \sin\theta \\
		0 & \sin\theta & \cos\theta \\
	\end{array} \!\! \right), \\
	(R_a \otimes I_d) = \left( \!\! \begin{array}{ccc}
		-1 & 0 & 0 \\
		0 & 1 & 0 \\
		0 & 0 & 1 \\
	\end{array} \!\! \right),
\end{eqnarray*}
where
\[ \cos \theta = \frac{N-3}{N-1}, \quad {\rm and} \quad \sin \theta = \frac{2\sqrt{N-2}}{N-1}. \]
Combining these operators, the search operator \eref{eq:U} is
\[ U = \left( \!\! \begin{array}{ccc}
	-\rmi\sin\phi & -\cos\phi\cos\theta & \cos\phi\sin\theta \\
	-\cos\phi & -\rmi\sin\phi\cos\theta & \rmi\sin\phi\sin\theta \\
	0 & \rme^{\rmi\phi} \sin\theta & \rme^{\rmi\phi} \cos\theta \\
\end{array} \!\! \right). \]
The (unnormalized) eigenvectors and corresponding eigenvalues of this are
\[ \psi_{\phi} = \left( \!\! \begin{array}{c}
	-\cot \frac{\theta}{2} \\
	-\cot \frac{\theta}{2} \\
	1 \\
\end{array} \!\! \right), \quad -\rme^{\rmi\phi} \]
and
\[ \psi_{\pm} = \left( \!\! \begin{array}{c}
	(1 - \rme^{-\rmi(\mp \sigma + \phi)}) \csc\theta \\
	(-\cos\theta + \rme^{-\rmi(\mp \sigma + \phi)}) \csc\theta \\
	1 \\
\end{array} \!\! \right), \quad \rme^{\pm \rmi\sigma} \]
where
\[ \cos\sigma = \frac{(1 + \cos\theta) \cos\phi}{2}, \quad \sin\sigma = \frac{\sqrt{4 - (1 + \cos\theta)^2 \cos^2\phi}}{2}. \]

It will be useful in our analysis to have the sum and difference of $\psi_+$ and $\psi_-$ because they evolve to each other up to an overall factor of $\rmi$ (\textit{i.e.}, up to a phase) in $\pi/2\sigma$ applications of the search operator $U$ \eref{eq:U}. That is,
\begin{equation}
	\label{eq:evo}
	U^{\pi/2\sigma} (\psi_+ \pm \psi_-) = (\rme^{\rmi\sigma})^{\pi/2\sigma} \psi_+ \pm (-\rme^{\rmi\sigma})^{\pi/2\sigma} \psi_- = \rmi(\psi_+ \mp \psi_-).
\end{equation}
In particular, the sum and difference of $\psi_+$ and $\psi_-$ are
\begin{eqnarray*}
	\psi_+ + \psi_- = \left( \!\! \begin{array}{c}
		2(1 - \rme^{-\rmi\phi} \cos\sigma) \csc\theta \\
		2(-\cos\theta + \rme^{-\rmi\phi} \cos\sigma) \csc\theta \\
		2 \\
	\end{array} \!\! \right), \\
	\psi_+ - \psi_- = \left( \!\! \begin{array}{c}
		-2\rmi \rme^{-\rmi\phi} \sin\sigma \csc\theta \\
		2\rmi \rme^{-\rmi\phi} \sin\sigma \csc\theta \\
		0 \\
	\end{array} \!\! \right).
\end{eqnarray*}
Pluging in for $\sigma$ and $\theta$, they become
\begin{eqnarray*}
	\psi_+ + \psi_- = \left( \!\! \begin{array}{c}
		\frac{N - (N-2)\rme^{-2\rmi\phi}}{2\sqrt{N-2}} \\
		\frac{-(N-3) + \rme^{-\rmi\phi}(N-2)\cos\phi}{\sqrt{N-2}} \\
		2 \\
	\end{array} \!\! \right), \\
	\psi_+ - \psi_- = \left( \!\! \begin{array}{c}
		-\frac{\rmi \rme^{-\rmi\phi} \sqrt{(N-1)^2 - (N-2)^2 \cos^2\phi}}{\sqrt{N-2}} \\
		\frac{\rmi \rme^{-\rmi\phi} \sqrt{(N-1)^2 - (N-2)^2 \cos^2\phi}}{\sqrt{N-2}} \\
		0 \\
	\end{array} \!\! \right).
\end{eqnarray*}
We can use these sum and difference formulas to find the evolution of the search algorithm when $\phi$ scales less than, equal to, or greater than $1/\sqrt{N}$. To simplify the calculation, we substitue $\phi = c/\sqrt{N}$ and equivalently consider when $c$ scales less than, equal to, or greater than a constant. Then Taylor expanding for large $N$, we get
\begin{eqnarray}
	\psi_+ + \psi_- \approx \left( \!\! \begin{array}{c}
		\rmi c + O\left( \frac{c^2 + 1}{\sqrt{N}} \right) \\
		-\rmi c + O\left( \frac{c^2 - 1}{\sqrt{N}} \right) \\
		2 \\
	\end{array} \!\! \right), \label{eq:sum} \\
	\psi_+ - \psi_- = \left( \!\! \begin{array}{c}
		-\rmi\sqrt{c^2 + 2} + O\left( \frac{c\sqrt{c^2 + 2}}{\sqrt{N}} \right) \\
		\rmi\sqrt{c^2 + 2} + O\left( \frac{c\sqrt{c^2 + 2}}{\sqrt{N}} \right) \\
		0 \\
	\end{array} \!\! \right). \label{eq:diff}
\end{eqnarray}
Now let us consider each scaling of $\phi = c/\sqrt{N}$ or $c$ as separate cases.

%-------------------------------------------------------------------------------

\begin{figure}
\begin{center}
	\includegraphics{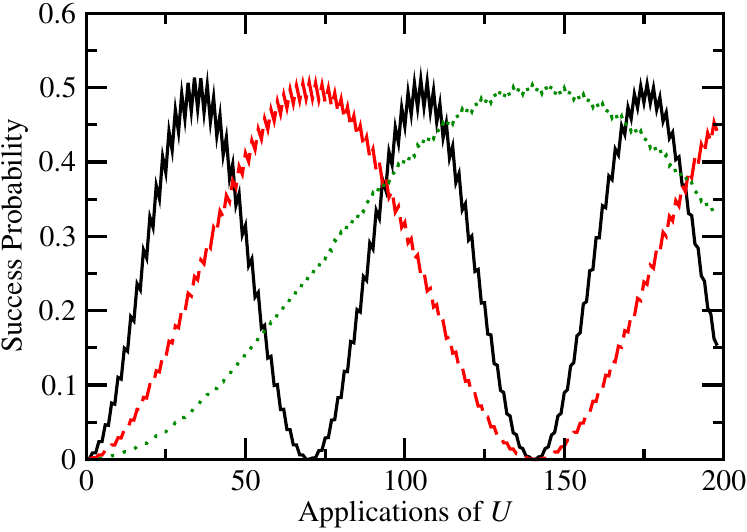}
	\caption{\label{fig:discrete_case1} Success probability as a function of the number of applications of $U$ for search on the complete graph with $\phi = 1/N^{3/4}$ and $N = 1024$, $4096$, and $16384$ vertices corresponding to the solid black, dashed red, and dotted green curves, respectively.}
\end{center}
\end{figure}

\textit{Case 1:} $\phi = o(1/\sqrt{N})$ or $c = o(1)$. When $\phi$ scales less than $1/\sqrt{N}$, or equivalently when $c$ scales less than a constant, the sum \eref{eq:sum} and difference \eref{eq:diff} of $\psi_+$ and $\psi_-$ have leading-order terms
\[ \psi_+ + \psi_- \approx \left( \!\! \begin{array}{c}
	0 \\
	0 \\
	2 \\
\end{array} \!\! \right), \quad
\psi_+ - \psi_- \approx \left( \!\! \begin{array}{c}
	-\rmi\sqrt{2} \\
	\rmi\sqrt{2} \\
	0 \\
\end{array} \!\! \right). \]
Then the initial state is approximately
\[ \ket{\psi_0} \approx \ket{bb} \approx \frac{1}{2} \left( \psi_+ + \psi_- \right). \]
Using \eref{eq:evo}, after $\pi/2\sigma$ applications of $U$, the system evolves to
\[ \frac{\rmi}{2} \left( \psi_+ - \psi_- \right) \approx \frac{1}{\sqrt{2}} \left( \!\! \begin{array}{c}
	1 \\
	-1 \\
	0 \\
\end{array} \!\! \right), \]
which is half in $\ket{ab}$ and half in $\ket{ba}$. This corresponds to a runtime of
\begin{eqnarray*}
	t_*
		&= \frac{\pi}{2\sigma} \approx \frac{\pi}{2\sin\sigma} = \frac{\pi}{\sqrt{4 - (1 + \cos\theta)^2 \cos^2\phi}} \approx \frac{\pi}{\sqrt{4\phi^2 + 8/N}} \\
		&\approx \frac{\pi}{2\sqrt{2}} \sqrt{N}.
\end{eqnarray*}
So if $\phi$ scales less than $1/\sqrt{N}$, the system evolves from $\ket{\psi_0} \approx \ket{bb}$ to being half in $\ket{ab}$ and half in $\ket{ba}$ after $\pi\sqrt{N}/2\sqrt{2}$ applications of $U$, which results in a success probability of $1/2$ when measuring the position of the particle due to the $\ket{ab}$ piece. This result is the same as without the potential barriers \cite{Wong10}, and so the potential barriers are too small to affect the search for large $N$. An example of this is shown in \fref{fig:discrete_case1} with $\phi = 1/N^{3/4}$ and $N = 1024$, $4096$, and $16384$. For each of these, the success probability reaches $1/2$ at respective times $\pi\sqrt{1024}/2\sqrt{2} \approx 36$, $\pi\sqrt{4096}/2\sqrt{2} \approx 71$, and $\pi\sqrt{16384}/2\sqrt{2} \approx 142$, as expected.

%-------------------------------------------------------------------------------

\begin{figure}
\begin{center}
	\includegraphics{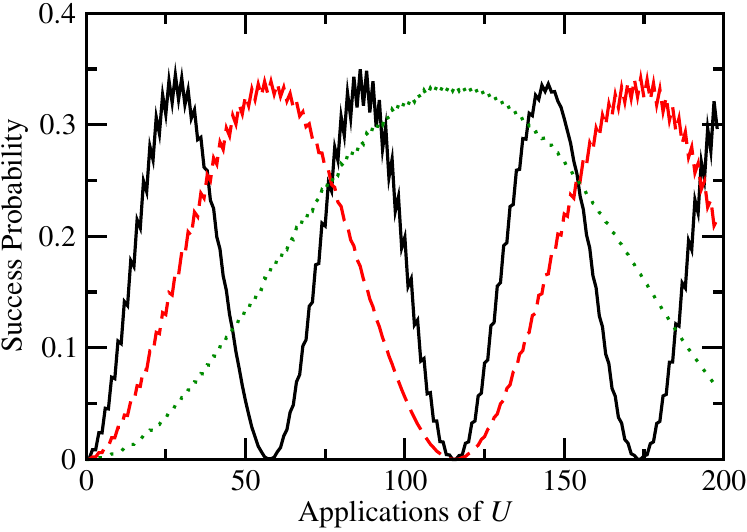}
	\caption{\label{fig:discrete_case2} Success probability as a function of the number of applications of $U$ for search on the complete graph with $\phi = 1/\sqrt{N}$ and $N = 1024$, $4096$, and $16384$ vertices corresponding to the solid black, dashed red, and dotted green curves, respectively.}
\end{center}
\end{figure}

\textit{Case 2:} $\phi = \Theta(1/\sqrt{N})$ or $c = \Theta(1)$. When $\phi = c/\sqrt{N}$ with coefficient $c$ constant, the sum \eref{eq:sum} and difference \eref{eq:diff} of $\psi_+$ and $\psi_-$ have leading-order terms
\[ \psi_+ + \psi_- \approx \left( \!\! \begin{array}{c}
	\rmi c \\
	-\rmi c \\
	2 \\
\end{array} \!\! \right), \quad
\psi_+ - \psi_- \approx \left( \!\! \begin{array}{c}
	-\rmi\sqrt{c^2+2} \\
	\rmi\sqrt{c^2+2} \\
	0 \\
\end{array} \!\! \right). \]
Then
\[ \left( \psi_+ + \psi_- \right) + \frac{c}{\sqrt{c^2+2}} \left( \psi_+ - \psi_- \right) \approx \left( \!\! \begin{array}{c} 0 \\ 0 \\ 2 \end{array} \!\! \right) = 2 \ket{bb}. \]
Since the initial state $\ket{\psi_0} \approx \ket{bb}$ is roughly half this, after $\pi/2\sigma$ applications of $U$, the state of the system is \eref{eq:evo}
\[ \frac{\rmi}{2} \left[ \left( \psi_+ - \psi_- \right) + \frac{c}{\sqrt{c^2+2}} \left( \psi_+ + \psi_- \right) \right] = \frac{\rmi}{2} \left( \!\! \begin{array}{c}
	\frac{-2\rmi}{\sqrt{c^2+2}} \\
	\frac{2\rmi}{\sqrt{c^2+2}} \\
	\frac{2c}{\sqrt{c^2+2}} \\
\end{array} \!\! \right). \]
Measuring the position of the particle yields a success probability given by the square of the first term, which corresponds to $\ket{ab}$:
\[ \left| \frac{\rmi}{2} \frac{-2\rmi}{\sqrt{c^2+2}} \right|^2 = \frac{1}{c^2+2}. \]
The corresponding runtime is
\begin{eqnarray*}
	t_*
	&= \frac{\pi}{2\sigma} \approx \frac{\pi}{2\sin\sigma} = \frac{\pi}{\sqrt{4\phi^2 + (3+\cos\theta)(1-\cos\theta)}} \approx \frac{\pi}{\sqrt{4\phi^2 + 8/N}} \\
	&= \frac{\pi}{2\sqrt{c^2+2}} \sqrt{N}.
\end{eqnarray*}
An example of this is shown in \fref{fig:discrete_case2} with $c = 1$ and $N = 1024$, $4096$, and $16384$. Each of these reach a success probability of $1/(1^2 + 2) = 1/3$ at respective times $\pi\sqrt{1024}/2\sqrt{1^2+2} \approx 29$, $\pi\sqrt{4096}/2\sqrt{1^2+2} \approx 58$, and $\pi\sqrt{16384}/2\sqrt{1^2+2} \approx 116$, as expected. While the number of steps and success probability are smaller than in the first case, their scalings are unchanged, so we still achieve Grover's $\Theta(\sqrt{N})$ steps with such potential barriers.

%-------------------------------------------------------------------------------

\begin{figure}
\begin{center}
	\includegraphics{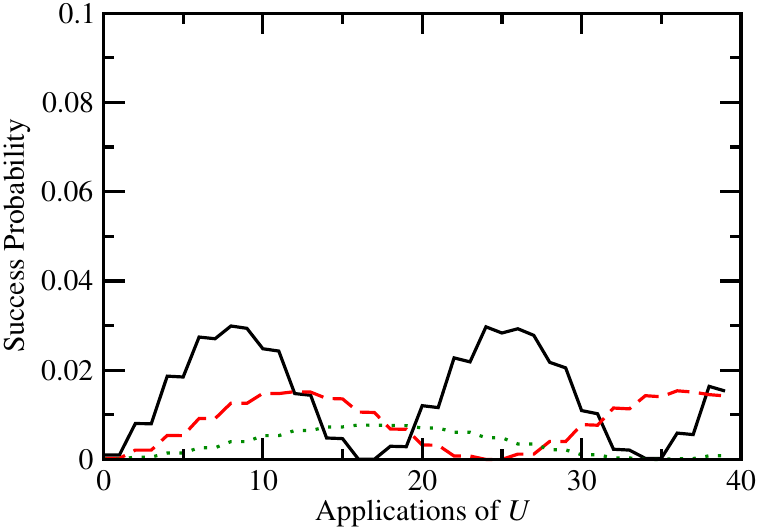}
	\caption{\label{fig:discrete_case3} Success probability as a function of the number of applications of $U$ for search on the complete graph with $\phi = 1/N^{1/4}$ and $N = 1024$, $4096$, and $16384$ vertices corresponding to the solid black, dashed red, and dotted green curves, respectively.}
\end{center}
\end{figure}

\textit{Case 3:} $\phi = \omega(1/\sqrt{N})$ or $c = \omega(1)$ (and $\phi = o(1)$ or $c = o(\sqrt{N})$). When $\phi$ scales larger than $1/\sqrt{N}$ (but still less than a constant), or equivalently when $c$ scales greater than a constant (but less than $\sqrt{N}$), the sum and difference of $\psi_+$ and $\psi_-$ for large $N$ are
\[ \psi_+ + \psi_- = \left( \!\! \begin{array}{c}
	\rmi\sqrt{N}\phi \\
	-\rmi\sqrt{N}\phi \\
	2 \\
\end{array} \!\! \right), \quad \psi_+ - \psi_- = \left( \!\! \begin{array}{c}
	-\rmi\sqrt{N}\phi \\
	\rmi\sqrt{N}\phi \\
	0 \\
\end{array} \!\! \right). \]
Then
\[ 2 \psi_+ = (\psi_+ + \psi_-) + (\psi_+ - \psi_-) \approx \left( \!\! \begin{array}{c} 0 \\ 0 \\ 2 \end{array} \!\! \right) = 2 \ket{bb} \]
Thus the system approximately begins in an eigenstate:
\[ \ket{\psi_0} \approx \ket{bb} \approx \psi_+, \]
so it fails to evolve (apart from a global, unobservable phase), and hence the search fails. That is, measuring the state at later time equates to measuring the initial equal superposition state, which gives the marked vertex with probability $1/N$, which is equivalent to classically guessing and checking. An example of this failure is shown in \fref{fig:discrete_case3} with $\phi = 1/N^{1/4}$. As $N$ increases, the success probability evolves less and less from its initial value of $1/N$.

%-------------------------------------------------------------------------------

\begin{figure}
\begin{center}
	\includegraphics{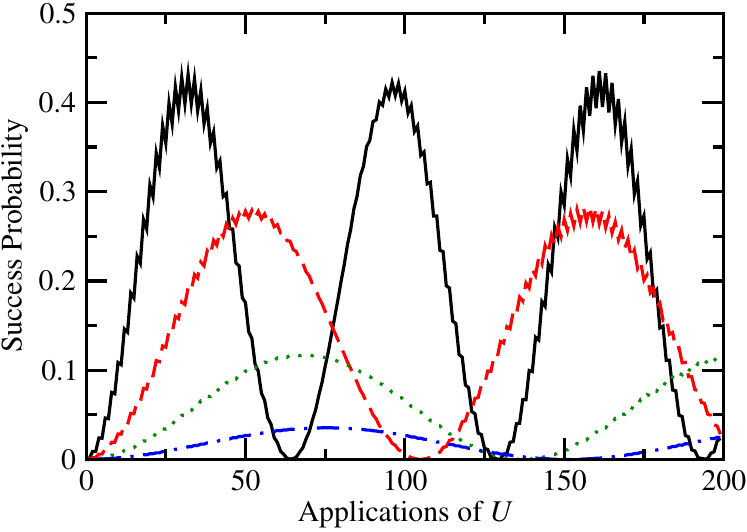}
	\caption{\label{fig:discrete_case4} Success probability as a function of the number of applications of $U$ for search on the complete graph with $\phi = 0.02$ and $N = 1024$, $4096$, $16384$, and $65536$ vertices corresponding to the solid black, dashed red, dotted green, and dot-dashed blue curves, respectively.}
\end{center}
\end{figure}

\textit{Case 4:} $\phi = \Theta(1)$ or $c = \Theta(\sqrt{N})$. When $\phi$ scales as a constant, which is the largest it can scale since $\phi = \pi/2$ corresponds to the potential barriers stopping all transitions, then the behavior from Case 3 persists---for large $N$, the system approximately begins in an eigenstate, and so the success probability does not evolve. This is shown in \fref{fig:discrete_case4}, where for constant $\phi$, increasing $N$ causes the success probability to evolve less and less.

%-------------------------------------------------------------------------------

Thus we see an abrupt change in the behavior of the algorithm for large $N$, depending on $\phi$: when $\phi = O(1/\sqrt{N})$, the full quantum quadratic speedup is achieved, and when $\phi = \omega(1/\sqrt{N})$, no speedup over classical is provided. This is summarized in \tref{table:summary}. Since the faulty shift is $\alpha S + \beta I$ with $\alpha = \cos\phi$ and $\beta = \rmi\sin\phi$, the amplitude of the particle failing to hop is $\beta = \rmi\sin\phi \approx \rmi \phi$ for small $\phi$. Therefore the failure amplitude $\beta$ must scale as $O(1/\sqrt{N})$ for search to retain its quantum speedup; otherwise, the algorithm searches in classical time. This means the failure amplitude must decrease as $N$ increases, so searching larger ``databases'' requires increasingly reliable hop operators in the absense of error correction. In follow-up work with Ambainis \cite{Wong15}, we show how the coin and oracle operators can be modified to partially offset the potential barriers, recovering the $O(\sqrt{N})$ runtime so long as $\phi$ does not approach $\pi/2$. To do so, we novelly interpret the quantum walk algorithm as an amplitude amplification algorithm, then adjust the phases applied to boost the success probability \cite{Long1999,Hoyer2000}.

\begin{table}
	\caption{\label{table:summary} Summary of discrete-time quantum walk search on the complete graph with potential barriers for large $N$. In the last case, the system stays in its initial uniform state, so the quantum algorithm is equivalent to classically guessing and checking.}
	\begin{indented}
	\item[]\begin{tabular}{@{}cccc}
		\br
		\textbf{Barrier} & \textbf{Runtime} & \textbf{Success Probability} & \textbf{Example(s)} \\
		\mr
		$\phi = o(1/\sqrt{N})$ & $\frac{\pi}{2\sqrt{2}} \sqrt{N}$ & $\frac{1}{2}$ & \Fref{fig:discrete_case1} \\[2ex]
		$\phi = c/\sqrt{N}$, constant $c$ & $\frac{\pi}{2\sqrt{c^2+2}} \sqrt{N}$ & $\frac{1}{c^2+2}$ & \Fref{fig:discrete_case2} \\[2ex]
		$\phi = \omega(1/\sqrt{N})$ & Not Applicable & $\frac{1}{N}$ & Figures~\ref{fig:discrete_case3} and \ref{fig:discrete_case4} \\
		\br
	\end{tabular}
	\end{indented}
\end{table}

%-------------------------------------------------------------------------------
% Section
%-------------------------------------------------------------------------------

\section{Continuous-Time Quantum Walks}

We end by discussing continuous-time quantum walks, which do not require the additional ``coin'' space, so the system evolves in the vertex Hilbert space $\mathbb{C}^N$. The system begins in the equal superposition over the vertices $\ket{s_v}$, and without potential barriers, evolves by Schr\"odinger's equation with Hamiltonian
\[ H = -\gamma A - \ketbra{a}{a}, \]
where $\gamma$ is an adjustable parameter corresponding to the jumping rate (amplitude per time), and $A$ is the adjacency matrix of the graph ($A_{ij} = 1$ if vertices $i$ and $j$ are adjacent). For the complete graph, all the vertices are connected to each other, so the adjacency matrix is
\[ A = \left( \!\! \begin{array}{cccc}
	0 & 1 & \dots & 1 \\
	1 & 0 & \dots & 1 \\
	\vdots & \vdots & \ddots & \vdots \\
	1 & 1 & \dots & 0 \\
\end{array} \!\! \right), \]
and it effects the quantum walk \cite{CG2004}. Note since the complete graph is regular, using the adjacency matrix is equivalent to using the graph Laplacian \cite{Wong19}. As shown in \fref{fig:complete}, there are only two types of vertices $a$ and $b$, so we can write $H$ in the 2D subspace spanned by $\ket{a}$ and $\ket{b} = \sum_{x \ne a} \ket{x} / \sqrt{N-1}$:
\[
	H = -\gamma \left( \!\! \begin{array}{cc}
		\frac{1}{\gamma} & \sqrt{N-1} \\
		\sqrt{N-1} & N-2 \\
	\end{array} \!\! \right).
\]
As shown in \cite{Wong6}, this has eigenstates
\[ \ket{\psi_{0,1}} \propto \ket{s_v} + \frac{1 - \gamma N \pm \Delta E}{2\gamma\sqrt{N}} \ket{a} \]
with gap in the corresponding eigenvalues $E_0$ and $E_1$
\[ \Delta E = E_1 - E_0 = \sqrt{(1 - \gamma N)^2 + 4\gamma}. \]
When $\gamma N$ takes its critical value of $1$, the energy gap is $\Delta E = 2/\sqrt{N}$ with eigenstates $\ket{\psi_{0,1}} \propto \ket{s_v} \pm \ket{a}$, so the system evolves from $\ket{s_v}$ to $\ket{a}$ with probability $1$ in time $\pi/\Delta E = \pi \sqrt{N}/2$ \cite{CG2004,Wong6,Wong10}. This behavior is retained near the critical $\gamma$ when $\gamma N = 1 + o(1/\sqrt{N})$ for large $N$. When $\gamma N = 1 + c/N$ for constant $c$, we get $\Delta E = \sqrt{(c^2+4)/N}$ and $\ket{\psi_{0,1}} \propto \ket{s_v} + (-c \pm \sqrt{c^2+4})/2 \ket{a} = \ket{s_v} + o(1) \ket{a}$ for large $N$, so the system evolves from $\ket{s_v}$ to $\ket{a}$ with constant probability in time $\pi\sqrt{N}/\sqrt{c^2+4}$. Finally, when $\gamma N = 1 + \omega(1/\sqrt{N})$, we get that $\ket{\psi_0}$ or $\ket{\psi_1}$ equals $\ket{s_v}$ for large $N$, depending on if the deviation is positive or negative, respectively. So the system begins in an eigenstate and fails to evolve beyond acquiring an unobservable phase, which is consistent with degenerate perturbation theory \cite{Wong5}. Thus we require $\gamma N = 1 + O(1/\sqrt{N})$ to achieve a $\Theta(\sqrt{N})$ search, and otherwise the search is no better than classically guessing and checking.

\begin{figure}
\begin{center}
	\includegraphics{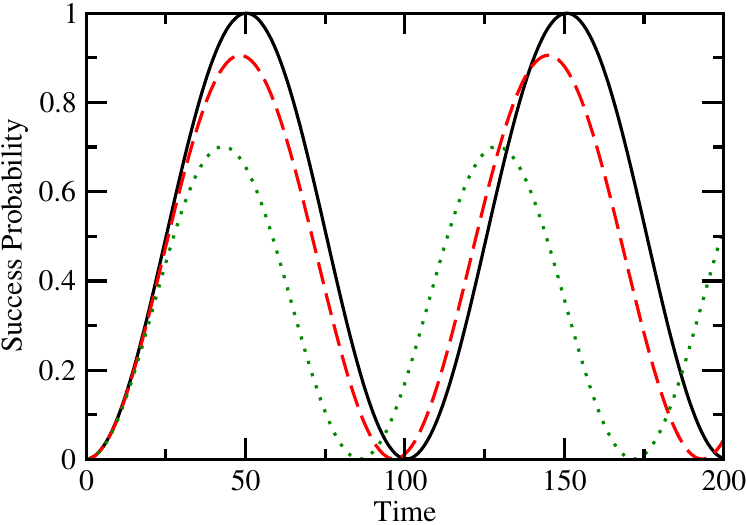}
	\caption{\label{fig:continuous} Success probability as a function of time for continuous-time search on the complete graph with $N=1024$ vertices, $\gamma = 1/N$, and $\epsilon = 0$ (no potential barriers), $\epsilon = 0.02$, and $\epsilon = 0.04$ corresponding to the solid black, dashed red, and dotted green curves, respectively.}
\end{center}
\end{figure}

Now with potential barriers, the amplitude that the particle hops from one vertex to another is decreased, say by $\epsilon$, and the amplitude of doing nothing is increased by the same amount. This effectively modifies the adjacency matrix to be
\[ A' = \left( \!\! \begin{array}{cccc}
	(N-1)\epsilon & 1-\epsilon & \dots & 1-\epsilon \\
	1-\epsilon & (N-1)\epsilon & \dots & 1-\epsilon \\
	\vdots & \vdots & \ddots & \vdots \\
	1-\epsilon & 1-\epsilon & \dots & (N-1)\epsilon \\
\end{array} \!\! \right). \]
But this is simply $(N-1)\epsilon I + (1-\epsilon)A$, and since the first term is a multiple of the identity matrix, it yields no observable effect and can be dropped. Thus the search Hamiltonian is effectively
\[ H = -\gamma (1-\epsilon) A - \ketbra{a}{a}. \]
The actual critical $\gamma$ for this walk solves $\gamma N (1-\epsilon) = 1$, which implies $\gamma N = 1/(1-\epsilon) \approx 1 + \epsilon$ for $\epsilon = o(1)$. If $\gamma N$ is still chosen to be its barrier-free value of $1$, then we are $\epsilon$ away from the true value. But above, we showed search in $\Theta(\sqrt{N})$ time is still possible when $\epsilon = O(1/\sqrt{N})$, which is the same constraint as the discrete-time quantum walk. That the system evolves less and less as the error $\epsilon$ grows is illustrated in \fref{fig:continuous}.

%-------------------------------------------------------------------------------
% Section
%-------------------------------------------------------------------------------

\section{Conclusion}

We have shown the effect of potential barriers hindering a randomly walking quantum particle from searching on the complete graph of $N$ vertices. In discrete-time, the amplitude of not hopping must scale less than or equal to $1/\sqrt{N}$ for the search to achieve Grover's $\Theta(\sqrt{N})$ runtime without error correction. Otherwise, no improvement over classical is achieved. Similar behavior holds for search by continuous-time quantum walk. So even though a quantum walk on the one-dimensional line retains its characteristic ballistic dispersion with potential barriers \cite{Wong17}, our result indicates that the barriers can have significant algorithmic consequences. Further research includes how the barriers affect other algorithms based on quantum walks, and the effects of non-uniform or random barriers.

%-------------------------------------------------------------------------------
% Acknowledgments
%-------------------------------------------------------------------------------

\ack
This work was supported by the European Union Seventh Framework Programme (FP7/2007-2013) under the QALGO (Grant Agreement No.~600700) project, and the ERC Advanced Grant MQC.

%-------------------------------------------------------------------------------
% References
%-------------------------------------------------------------------------------

\section*{References}
\bibliographystyle{iop}
\bibliography{refs}

\end{document}